**RESEARCH** **Open Access**

# Mapping knowledge translation and innovation processes in Cancer Drug Development: the case of liposomal doxorubicin

David Fajardo-Ortiz[1], Luis Duran[2], Laura Moreno[2], Hector Ochoa[3] and Victor M Castaño[4,5*]

**Abstract**

We explored how the knowledge translation and innovation processes are structured when theyresult in innovations, as in the case of liposomal doxorubicin research. In order to map the processes, a literature network analysis was made through Cytoscape and semantic analysis was performed by GOPubmed which is based in the controlled vocabularies MeSH (Medical Subject Headings) and GO (Gene Ontology). We found clusters related to different stages of the technological development (invention, innovation and imitation) and the knowledge translation process (preclinical, translational and clinical research), and we were able to map the historic emergence of Doxil as a paradigmatic nanodrug. This research could be a powerful methodological tool for decision-making and innovation management in drug delivery research.

## Introduction

Technological Innovation (TI) -the processes leading to the emergence of new technologies in the market- and Knowledge Translation (KT), –the conversion of research results into better practices- notoriously converge, more obvious than in any other scientific discipline, in biomedical sciences. The epistemological fundamental idea behind both concepts is that Science must serve, primordially, to enhance the life conditions of Humankind. Ever since the Enlightenment, modern history of human society cannot be understood without that powerful dream -Science helping to build a better world-, which has been repeated over and over again by scientists and philosophers, from Thomas Jefferson to Bertrand Russell [1,2]. Regardless the specific epistemological approach to Science, the transition from research results to the delivery of solutions to solve the needs of human society is the common base of innovation and knowledge translation.

However, both processes are based on two very different tautologies: KT is founded on a Medical and Health sciences, with the moral obligation to provide useful advice to build a better general health status [3], whereas TI aims to produce temporal monopolies in a competitive market context [4]. KT, therefore, is an effort to articulate basic research with clinical practice and health social goals and, to that effect, there is a continuing complaint about science and practices being, in practice, poorly communicated, i.e., that "we are lost in translation" [5].

In this regard, the United States government [6], as well as other countries [7], have made important, and costly, efforts to institutionalize strategies in order to accelerate KT. Additionally, models [8] and experiences [9] on how to close the gap between research and practices can be found more and more frequently in the specialized literature.

Accordingly, for the purposes of this article, TI process is understood as the entire process of technological change. TI process is composed by three phases: invention, innovation and imitation. Invention is the stage in which the technological base is created. Innovation starts when the final inventions are published and ends when the product is approved and delivered to the market. Imitation is the research and development of the international community following the success of the innovation leader [10]. In turn, KT is divided in three steps: first, when basic

* Correspondence: meneses@unam.mx
[4]Centro de Fisica Aplicada y Tecnologia Avanzada, Universidad Nacional Autonoma de Mexico, Queretaro, Mexico
[5]CIATEQ, Queretaro, Mexico
Full list of author information is available at the end of the article research is translated into clinical knowledge; second, when the latter is translated to clinical practice and finally, when these results in public health outcome [11].

It has been pointed out by various authors that there is a disproportion between the number of papers published and the amount of nanodrugs readily available in the market [10]. Indeed, there exist thousands of papers as compared to just 247 confirmed commercial products in a preclinical, clinical or commercial stage [12]. The question then arises on how TI is structured when a nanodrug is approved, that is, when an innovation truly emerges. The approach we offer herein to answer that question consists in mapping the aforementioned processes in the scientific literature network for a case example, namely liposomal doxorubicin. Liposomes are micro- and nano-scaled lipid bilayered "bubbles" that can be used in a plethora of therapeutic strategies. The therapeutic use of liposomes ranges from drug carriers and drug delivery systems to multiplex with antibodies, optical contrast, genetic material and others [13]. Liposomal doxorubicin was chosen as an illustrative case study, with a historical relevance, since a PEGylated liposomal doxorubicin, Doxil, was the first nanodrug approved by the United States Food and Drug Administration (FDA) in 1995 [14]. The proposed approach aims to understand how scientific research evolves and gets re-organized when a technological change process is taking place. Moreover, this approach is aimed to provide insight towards building more effective innovation policies in health-oriented nanotechnologies.

This paper is based on a strong methodological and exploratory approach. In this research, we identified the concurrence of the TI and KT stages in the literature network of liposomal doxorubicin research. The knowledge translation pipeline in nanomedicine is a time-consuming and complex road that requires acceleration [15]. This methodology could be useful for the development of an innovation roadmap for nanomedicine. This mapping strategy could inform about which research groups, ideas or approaches are effectively connecting the basic research with the clinical observation [16,7]. Moreover, we have shown, in a previous research, how this methodology could provide accurate information about the current status of development of a nanomedicine research. For example, we have previously reported that two different types of cancer nanotechnologies (liposomes or metallic nanoparticles) are, actually, in two different stages of KT and TI [17].

Accordingly, three different research strategies have been developed to analyze KT in literature networks. The first consists of classifying the papers either as discovery (of risk factor associated with disease) or delivery (of the interventions) research and identifying the key cross-citation between these research fields [18,19]. The second strategy consists in classifying the entire journals as basic research, clinical research, clinical mix or clinical observation, according the distribution of terms occurring in the titles of the articles published in each journal. The journals inter-citations are then analyzed [20,21]. Finally, we have developed an innovative methodology [16,17] which combines semantic analysis with network analysis to identify hidden colleges and/or stages of the innovation process. This methodology is based on the semantic analysis performed by GOPUMED [22], which is based in controlled vocabularies (Medical Subject Heading [23] and Gene Ontology [24]).

## Material and methods

A search of research papers on liposomal doxorubicin (TITLE: (liposom*) AND TOPIC:(doxorubicin); Timespan = All years; Indexes = SCI-EXPANDED, SSCI, A&HCI, CPCI-S, CPCI-SSH) was made in the Web of Science (WoS) [25] as for February 2014.

Twenty percent of the most cited papers were selected. The distribution of citations in the scientific literature follows a power law [26] i.e., there are few papers with a huge quantity of citations and there are thousands of manuscripts with few or no citations. Therefore, it is meaningless to speak about representative samples. Moreover, power laws could be simplified with a 20/80 rule, i.e., 20% of papers receive 80% of the citations [27]. When we consulted the WoS for this particular field (liposomes and doxorubicin) we found that this rule is fairly close to what it is reported: 20% of the most cited papers received 83% of the citations. Therefore, we reasonably chose the 20% of the most cited papers because this quantity is large enough to get most of the communication process that happen in the literature network.

The software packages, Hiscite [28] and Pajek [29], were used to build the citation network model. Cytoscape [30] software was used to visualize and analyze the network model. Clust&See [31], a Cytoscape plug-in, was used to identify clusters (Tfit algorithm: modularity criterion, multilevel transfer optimization), which could be related to different stages on KT evolution or invisible colleges. The Clust&See algorithms are based on the optimization of Newman's modularity [31]. This approach define clusters as "groups of vertices within which connections are dense but between which they are sparse" [32]. Therefore, this algorithms put together papers that have the same citation (connection) profile. In addition, Clust&See represented each found cluster as a single meta-node with a size width is proportional to the number of their constituent nodes. Similarly, the meta-nodes are connected by meta-edges, whose width is proportional to the number of interactions among their vertices.

The papers of the network model were searched in the engine GOPubMed [22], which semantically analyzed them by attaching each paper to terms from Medical Subject Heading (MeSH) and Gene Ontology (GO). We calculated the rate of clinical terms vs. non-clinical, since we had previously defined as clinical terms all MeSH terms belonging to the next higher hierarchy categories: "Diagnosis", "Therapeutics", "Surgical Procedures, Operative", "Named Groups", "Persons" and "Health Care".

The most central papers of each cluster (by higher hierarchy and effective degree) were identified. Literature networks constitute a particular type of genealogical graphs [28,29], i.e., they are unidirectional networks in which the papers have ancestors and/or descendants. The paper with the highest hierarchy is the common ancestor of the most of manuscripts in a cluster (subnetwork). Effective (weighted) degree is a measure of centrality. This measure is calculated by counting the effective number of edge weights connected to the given node [33], i.e. how much a node is connected to the most connected nodes.

We labeled the nodes (papers) of the network model with the name of the institution and the country of the correspondence author. In addition, we identify the main institutions of each cluster by their number of papers. This information could be useful to understand the relevance of the international collaboration for the liposomal doxorubicin innovation process.

The network model layout was displayed using the "spring embedded" algorithm of Cytoscape [30]. Where nodes act like particles that repels each others and connections that act as springs. The resulting layout (with the minimal sum of force in the network) puts together the papers that tend to cite the same documents and separates the papers that differ in their citing profile. The nodes were colored according a continuous scale (from red to blue), which is a function of the clinical terms rate.

## Results

1,747 papers related to liposomal doxorubicin were found in WoS. 20% of those documents (350) were selected to build the network model. These 350 papers received 30,360 citations (without counting auto-citations), which represents 83% of the total citations received by the 1,747 documents found. This overwhelming percentage of citations provides a clue on the importance of this 20% of the documents to the scientific communication on liposomal doxorubicin through the literature network.

342 of these papers form one single citation weak network, i.e., a network of papers connected for a least one citation. We made sure that the each document are labeled by GOPUBMED with the MeSH terms "doxorubicin" and "liposome", and discarded the papers that did not have these terms. Finally, a literature network of 274 papers was obtained (Figure 1).

Through the citation patrons of the network, Clust&see identified 8 clusters of highly interconnected papers (nodes) with the Tfit algorithm (Figure 1). After the identification of the cluster, we performed an analysis of the distribution of MesH and Go terms for each cluster. This distribution suggests that clusters correspond to different stages of TI and KT processes. Cluster 1 is conformed by basic research papers and it is related to a invention stage. Cluster 4 is basic research whereas cluster 2 papers are clinical researches, both clusters constitute the innovation stage, and the communication between them represents the knowledge translation process. Cluster 3 and 5 papers are basic researches and they represent a sort of incremental innovation of liposomal doxorubicin technology. Finally, Cluster 6, 7 and 8 are clinical observation researches aimed to extend the usage of this nanodrug to others types of cancer (Figure 1 and Tables 1 and 2).

Clusters are next described in a chronological order, according to the average year of publication of their papers. We identified the clusters by their size rank (number of documents).

Cluster 1, the oldest, is a subnetwork of 61 papers published between 1981 and 2009, being 1994 the average year. Most of papers of cluster 1 correspond to basic research. The average rate of clinical terms is 0.09, i.e., the papers of this cluster correspond to basic research. MeSH and Go terms are common to the others clusters, i.e., there is no specialization in this subnetwork. The papers with more effective degree and more centrality values are about liposome size, drug charging, and lowering the cardiotoxicity (Table 2). The basic research profile of this cluster, the average publishing year, its low density, the topics of the most central papers, and the non-specialized terms distributions suggest that cluster 1 is related to invention stage of liposomal doxorubicin.

Clusters 4 and 2, the next in the chronological order, are built by 35 and 57 documents, respectively. The cluster 4 papers were published between 1991 and 2009, and 1997 is the average year. Cluster 2 papers were published between 1994 and 2007, with 2000 being the average year. Clusters 4 and 2 have the strongest interconnectivity through 149 citations (Figure 2). Both clusters share the distinctive MeSH terms "Terapeutics" and "Drug carriers" but the cluster 4 papers includes animal research (mice), and cluster 2 is focused on patient treatment (Table 1). The clinical terms rate of cluster 4 is 0.11 which indicates that the papers are mostly basic research. Since the rate for cluster 2 is 0.278, these papers trend to be translational clinical research. The most central papers (by hierarchy and effective degree) of both clusters are about Doxil, i.e., PEGylated liposomal doxorubicin (Table 2). All of above suggests the TI process and KT take place in the interaction between clusters 2 and 4.

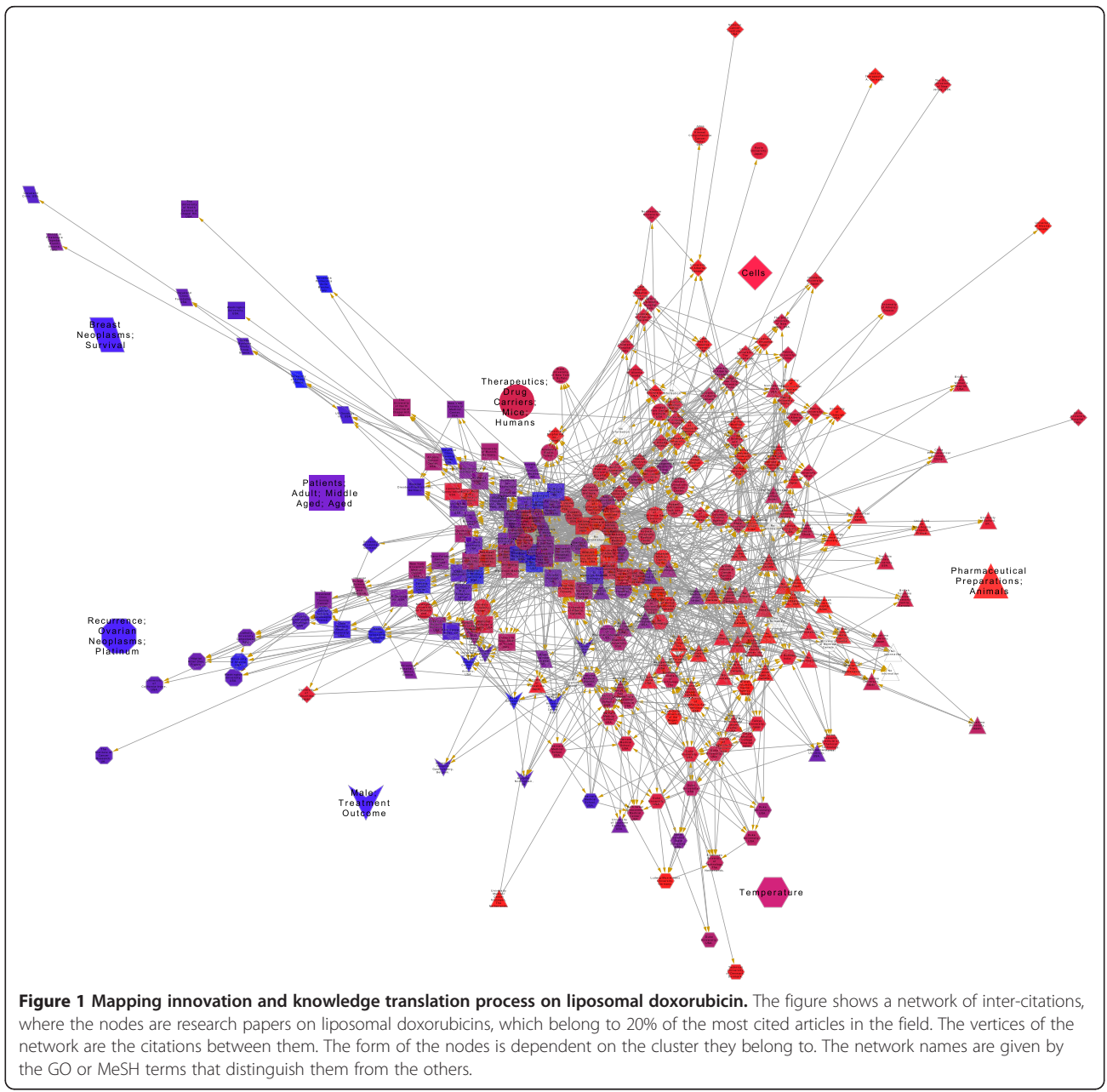

**Figure 1 Mapping innovation and knowledge translation process on liposomal doxorubicin.** The figure shows a network of inter-citations, where the nodes are research papers on liposomal doxorubicins, which belong to 20% of the most cited articles in the field. The vertices of the network are the citations between them. The form of the nodes is dependent on the cluster they belong to. The network names are given by the GO or MeSH terms that distinguish them from the others.

Clusters 5 and 3 are mainly connected to cluster 4 (Figure 1), and the average years are 2001 and 2003, respectively. Both clusters correspond to basic research. The corresponding MeSH and GO terms for cluster 3 are "Cells" and "cell" (Table 1). The central papers of cluster 3 are about targeting the cancerous cells (Table 2). Cluster 5 distinctive MeSH term is "temperature" (Table 1). The central papers of cluster 5 are about hyperthermia as an anti-cancer therapy (Table 2). The features of clusters 3 and 5 suggest that both constitute a sort of creative bifurcation of the main innovation timeline of liposomal doxorubicin, aimed to improve or to amplify its therapeutic performance.

Finally, clusters 6, 7, and 8 are small groups of clinical observation papers, with an average rate of clinical terms of 0.36, 0.39 and 0.41, respectively. The three cluster are extensions of cluster 2 (Figure 1 and 2). These clusters are related to a imitation phase which attempts to improve the efficiency, combine with other treatments, and extend the Doxil application to different types of cancer (Tables 1 and 2).

Finally, the correspondence information of each paper (Figure 1 and Table 3) shows that the leading research institutions in liposomal doxorubicin are mainly located in the United States, Canada, and Israel. The leading institution along the innovation history of the liposomal doxorubicin is the Hadassah Medical Organization (Israel).

**Table 1 Top MeSh and GO terms associated with papers belonging to each of the eight clusters**

| MeSH and GO terms | Papers labeled with the term |
|---|---|
| Cluster 1 | |
| Doxorubicin | 58 |
| Liposomes | 58 |
| Pharmaceutical preparations | 45 |
| Animals | 38 |
| Neoplasms | 32 |
| Cluster 2 | |
| Doxorubicin | 57 |
| Liposomes | 57 |
| Humans | 50 |
| Patients | 44 |
| Female | 44 |
| Neoplasms | 41 |
| Adult | 35 |
| Middle aged | 35 |
| Therapeutics | 30 |
| Aged | 29 |
| Drug carriers | 29 |
| Cluster 3 | |
| Liposomes | 50 |
| Doxorubicin | 48 |
| Cells | 40 |
| cell | 40 |
| Animals | 38 |
| Pharmaceutical preparations | 37 |
| Mice | 36 |
| Humans | 35 |
| Neoplasms | 31 |
| Cluster 4 | |
| Liposomes | 33 |
| Doxorubicin | 30 |
| Pharmaceutical preparations | 30 |
| Neoplasms | 27 |
| Animals | 27 |
| Therapeutics | 24 |
| Drug Carriers | 22 |
| Mice | 22 |
| Humans | 19 |
| Cluster 5 | |
| Doxorubicin | 34 |
| Liposomes | 34 |
| Pharmaceutical preparations | 28 |
| Animals | 25 |
| Neoplasms | 25 |
| Temperature | 18 |
| Cluster 6 | |
| Doxorubicin | 16 |
| Humans | 16 |
| Liposomes | 15 |
| Patients | 14 |
| Male | 14 |
| Middle aged | 12 |
| Aged | 12 |
| Female | 12 |
| Treatment outcome | 11 |
| Polyethylene glycols | 9 |
| Therapeutics | 9 |
| Adult | 9 |
| Cluster 7 | |
| Doxorubicin | 10 |
| Liposomes | 10 |
| Patients | 9 |
| Humans | 9 |
| Female | 8 |
| Breast neoplasms | 8 |
| Survival | 7 |
| Neoplasms | 7 |
| Therapeutics | 7 |
| Aged | 7 |
| Adult | 6 |
| Middle aged | 6 |
| Safety | 6 |
| Cluster 8 | |
| Doxorubicin | 9 |
| Liposomes | 9 |
| Patients | 9 |
| Recurrence | 9 |
| Survival | 9 |
| Adult | 9 |
| Aged | 9 |
| Female | 9 |
| Humans | 9 |
| Middle aged | 9 |
| Ovarian neoplasms | 9 |
| Platinum | 8 |
| Polyethylene glycols | 8 |
| Disease-free survival | 7 |
| Neoplasms | 7 |
| Safety | 6 |
| Aged, 80 and over | 6 |

**Table 2 Manuscripts with the highest degree of hierarchy within each of the clusters**

| Position | Authors | Citation | Title |
| --- | --- | --- | --- |
| Cluster 1 | | | |
| The first by degree | Mayer, L D, Tai, L C, Ko, D S, Masin, D, Ginsberg, R S, Cullis, P R, Bally, M B | Cancer research (Cancer Res), Vol. 49 (21): 5922–30, 1989 | Influence of vesicle size, lipid composition, and drug-to-lipid ratio on the biological activity of liposomal doxorubicin in mice. |
| Second by degree | Mayer, L D, Tai, L C, Bally, M B, Mitilenes, G N, Ginsberg, R S, Cullis, P R | Biochimica et biophysica acta (Biochim Biophys Acta), Vol. 1025 (2): 143–51, 1990 | Characterization of liposomal systems containing doxorubicin entrapped in response to pH gradients. |
| Second by degree | Herman, E H, Rahman, A, Ferrans, V J, Vick, J A, Schein, P S | Cancer research (Cancer Res), Vol. 43 (11): 5427–32, 1983 | Prevention of chronic doxorubicin cardiotoxicity in beagles by liposomal encapsulation. |
| The first by hierarchy | Forssen, E A, Tökès, Z A | Proceedings of the National Academy of Sciences of the United States of America (P Natl Acad Sci Usa), Vol. 78 (3): 1873–7, 1981 | Use of anionic liposomes for the reduction of chronic doxorubicin-induced cardiotoxicity. |
| Cluster 2 | | | |
| The first by degree | Muggia, F M, Hainsworth, J D, Jeffers, S, Miller, P, Groshen, S, Tan, M, Roman, L, Uziely, B, Muderspach, L, Garcia, A, Burnett, A, Greco, F A, Morrow, C P, Paradiso, L J, Liang, L J . | Journal of clinical oncology: official journal of the American Society of Clinical Oncology (J Clin Oncol), Vol. 15 (3): 987–93, 1997 | Phase II study of liposomal doxorubicin in refractory ovarian cancer: antitumor activity and toxicity modification by liposomal encapsulation. |
| Second by degree and first by hierarchy | Gabizon, A, Catane, R, Uziely, B, Kaufman, B, Safra, T, Cohen, R, Martin, F, Huang, A, Barenholz, Y | Cancer research (Cancer Res), Vol. 54 (4): 987–92, 1994 | Prolonged circulation time and enhanced accumulation in malignant exudates of doxorubicin encapsulated in polyethylene-glycol coated liposomes. |
| Second by degree | Uziely, B, Jeffers, S, Isacson, R, Kutsch, K, Wei-Tsao, D, Yehoshua, Z, Libson, E, Muggia, F M, Gabizon, A | Journal of clinical oncology : official journal of the American Society of Clinical Oncology (J Clin Oncol), Vol. 13 (7): 1777–85, 1995 | Liposomal doxorubicin: antitumor activity and unique toxicities during two complementary phase I studies. |
| Cluster 4 | | | |
| The first by degree and first by hierarchy | Papahadjopoulos, D, Allen, T M, Gabizon, A, Mayhew, E, Matthay, K, Huang, S K, Lee, K D, Woodle, M C, Lasic, D D, Redemann, C | Proceedings of the National Academy of Sciences of the United States of America (P Natl Acad Sci Usa), Vol. 88 (24): 11460–4, 1991 | Sterically stabilized liposomes: improvements in pharmacokinetics and antitumor therapeutic efficacy. |
| Second by degree | Vaage, J, Mayhew, E, Lasic, D, Martin, F | International journal of cancer. (Int J Cancer), Vol. 51 (6): 942–8, 1992 | Therapy of primary and metastatic mouse mammary carcinomas with doxorubicin encapsulated in long circulating liposomes. |
| Cluster 3 | | | |
| The first by degree | ElBayoumi, Tamer A, Torchilin, Vladimir P | Clinical cancer research : an official journal of the American Association for Cancer Research (Clin Cancer Res), Vol. 15 (6): 1973–80, 2009 | Tumor-targeted nanomedicines: enhanced antitumor efficacy in vivo of doxorubicin-loaded, long-circulating liposomes modified with cancer-specific monoclonal antibody. |
| Second by degree | Garde, Seema V, Forté, André J, Ge, Michael, Lepekhin, Eugene A, Panchal, Chandra J, Rabbani, Shafaat A, Wu, Jinzi J | Anti-cancer drugs (Anti-cancer Drug Des), Vol. 18 (10): 1189–200, 2007 | Binding and internalization of NGR-peptide-targeted liposomal doxorubicin (TVT-DOX) in CD13-expressing cells and its antitumor effects. |
| The first by hierarchy | Lee, R J, Low, P S | Biochimica et biophysica acta (Biochim Biophys Acta), Vol. 1233 (2): 134–44, 1995 | Folate-mediated tumor cell targeting of liposome-entrapped doxorubicin in vitro. |

**Table 2 Manuscripts with the highest degree of hierarchy within each of the clusters** (Continued)

| | | | |
|---|---|---|---|
| Cluster 5 | | | |
| The first by degree | Kong, G, Anyarambhatla, G, Petros, W P, Braun, R D, Colvin, O M, Needham, D, Dewhirst, M W | Cancer research (Cancer Res), Vol. 60 (24): 6950–7, 2000 | Efficacy of liposomes and hyperthermia in a human tumor xenograft model: importance of triggered drug release. |
| Second by degree | Needham, D, Anyarambhatla, G, Kong, G, Dewhirst, M W | Cancer research (Cancer Res), Vol. 60 (5): 1197–201, 2000 | A new temperature-sensitive liposome for use with mild hyperthermia: characterization and testing in a human tumor xenograft model. |
| The first by hierarchy | Mayhew, E G, Goldrosen, M H, Vaage, J, Rustum, Y M | Journal of the National Cancer Institute (J Natl Cancer I), Vol. 78 (4): 707–13, 1987 | Effects of liposome-entrapped doxorubicin on liver metastases of mouse colon carcinomas 26 and 38. |
| Cluster 6 | | | |
| The first by degree and first by hierarchy | Northfelt, D W, Martin, F J, Working, P, Volberding, P A, Russell, J, Newman, M, Amantea, M A, Kaplan, L D | Journal of clinical pharmacology (J Clin Pharmacol), Vol. 36 (1): 55–63, 1996 | Doxorubicin encapsulated in liposomes containing surface-bound polyethylene glycol: pharmacokinetics, tumor localization, and safety in patients with AIDS-related Kaposi's sarcoma. |
| Second by degree | Hussein, Mohamad A, Wood, Laura, Hsi, Eric, Srkalovic, Gordan, Karam, MaryAnn, Elson, Paul, Bukowski, Ronald M | Cancer, Vol. 95 (10): 2160–8, 2002 | A Phase II trial of pegylated liposomal doxorubicin, vincristine, and reduced-dose dexamethasone combination therapy in newly diagnosed multiple myeloma patients. |
| Cluster 7 | | | |
| The first by degree | Batist, G, Ramakrishnan, G, Rao, C S, Chandrasekharan, A, Gutheil, J, Guthrie, T, Shah, P, Khojasteh, A, Nair, M K, Hoelzer, K, Tkaczuk, K, Park, Y C, Lee, L W. | Journal of clinical oncology: official journal of the American Society of Clinical Oncology (J Clin Oncol), Vol. 19 (5): 1444–54, 2001 | Reduced cardiotoxicity and preserved antitumor efficacy of liposome-encapsulated doxorubicin and cyclophosphamide compared with conventional doxorubicin and cyclophosphamide in a randomized, multicenter trial of metastatic breast cancer. |
| Second by degree | Harris, Lyndsay, Batist, Gerald, Belt, Robert, Rovira, Douglas, Navari, Rudolph, Azarnia, Nozar, Welles, Lauri, Winer, Eric, TLC D-99 Study Group. | Cancer, Vol. 94 (1): 25–36, 2002 | Liposome-encapsulated doxorubicin compared with conventional doxorubicin in a randomized multicenter trial as first-line therapy of metastatic breast carcinoma. |
| The first by hierarchy | Balazsovits JA, Mayer LD, Bally MB, Cullis PR, McDonell M, Ginsberg RS, Falk RE. | Cancer Chemother Pharmacol. 1989;23(2):81–6. | Analysis of the effect of liposome encapsulation on the vesicant properties, acute and cardiac toxicities, and antitumor efficacy of doxorubicin. |
| Cluster 8 | | | |
| The first by degree and first by hierarchy | Gordon AN, Fleagle JT, Guthrie D, Parkin DE, Gore ME, Lacave AJ. | J Clin Oncol. 2001 Jul 15;19(14):3312–22. | Recurrent epithelial ovarian carcinoma: a randomized phase III study of pegylated liposomal doxorubicin versus topotecan. |
| Second by degree | Gordon, Alan N, Tonda, Margaret, Sun, Steven, Rackoff, Wayne, Doxil Study 30–49 Investigators | Gynecologic oncology (Gynecol Oncol), Vol. 95 (1): 1–8, 2004 | Long-term survival advantage for women treated with pegylated liposomal doxorubicin compared with topotecan in a phase 3 randomized study of recurrent and refractory epithelial ovarian cancer |

Papers with the highest hierarchy are the common ancestors of the documents of each cluster. Effective degree is the number of weighted connections to other papers (give a idea about how much connected is a paper to others highly connected documents of the network). Sometimes, the same paper has the highest hierarchy and most effective degree in a cluster.

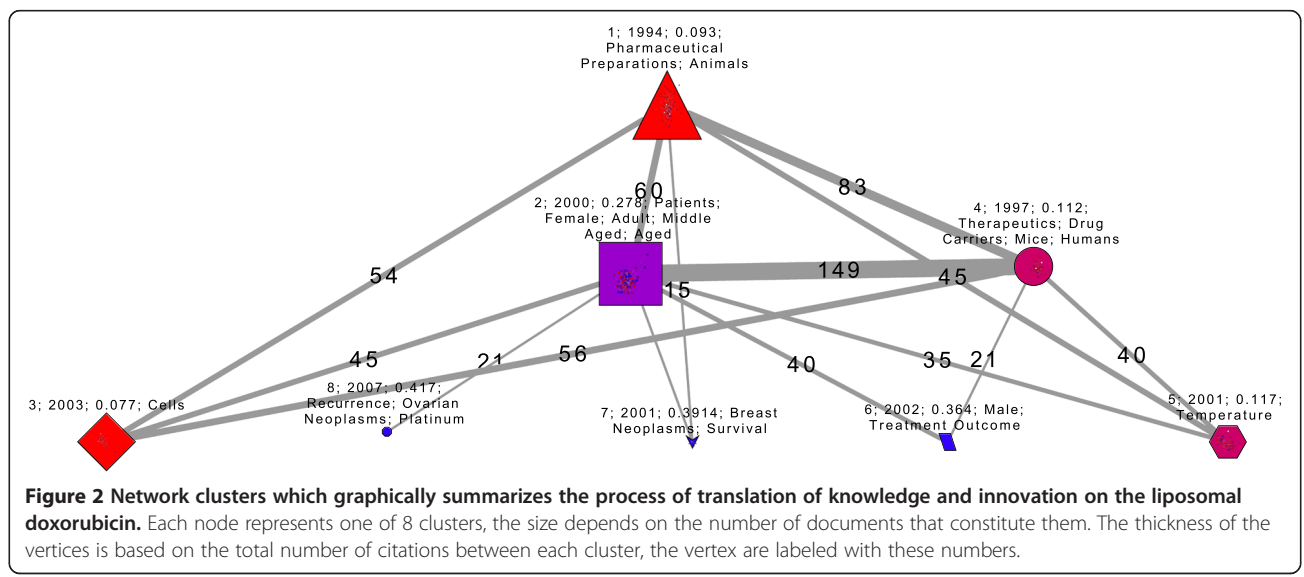

**Figure 2 Network clusters which graphically summarizes the process of translation of knowledge and innovation on the liposomal doxorubicin.** Each node represents one of 8 clusters, the size depends on the number of documents that constitute them. The thickness of the vertices is based on the total number of citations between each cluster, the vertex are labeled with these numbers.

Moreover, SEQUUS Pharmaceuticals, Inc. at Menlo Park (USA), Hadassah Medical Organization and the University of Alberta (Canada) which are institutions that participated in the innovation of Doxil [14] are the main actors in the cluster 2 (Table 3).

**Table 3 Main institution by their number of papers in each cluster**

|  |  |  |
|---|---|---|
| Cluster 1 | Hadassah Medical Organization, Israel. | 6 |
|  | British Columbia Cancer Agency, Canada. | 5 |
|  | Georgetown University, USA. | 5 |
|  | Hiroshima University, Japan. | 3 |
|  | University of British Columbia, Canada. | 3 |
|  | No information | 14 |
| Cluster 2 | University of Southern California, USA. | 5 |
|  | Hadassah Medical Organization, Israel. | 5 |
|  | New York University, USA. | 3 |
|  | SEQUUS Pharmaceuticals, Inc., Menlo Park, USA. | 3 |
|  | University of Alberta, Canada. | 3 |
| Cluster 3 | University of Alberta, Canada. | 10 |
|  | Hadassah Medical Organization, Israel. | 4 |
|  | Northeastern University, USA. | 3 |
|  | Peking University, China. | 3 |
| Cluster 4 | Hadassah Medical Organization, Israel. | 10 |
|  | Roswell Park Cancer Institute, USA. | 8 |
| Cluster 5 | Duke University, USA. | 7 |
|  | Harvard Medical School, USA. | 6 |
|  | University of California-San Francisco, USA | 3 |
| Cluster 6 | Cleveland Clinic Foundation, USA. | 3 |
| Cluster 7 and 8 | All institution have just one paper in theses clusters | 1 |

We only consider the corresponding address of the papers.

## Discussion

It is important to point out the meaning of our results. It has been pointed out that a network of highly cited papers is related to the paradigmatic structure of a specific research field [34]. A literature network could be considered as a interconnected set of information routes, then the topological position of a node, and therefore how many times a paper is cited, matters with regard to how much information it could actually transmit [35]. The network model that we built from the 20% most cited papers represents the core and the organizer of the communication process of the liposomal doxorubicin R&D. However, care must be taken not to confuse influence with veracity, validity or research quality: an influential paper is not always the best, in terms of scientific quality (whatever measure is used) [36,37].

Clustering methods are very useful to identify hidden colleges, that is, informal networks of researchers that read, cite, and interpret the reality of a problem in a very similar way. In addition, clustering could serve to identify paradigm shifts [38]. Clustering can map both informal colleges arrangements and the paradigm shift, because the former emerge from the latter. What we are seeing in our maps is the emergence of Doxil as the paradigmatic model of liposomal drugs (Figures 1 and 2). As mentioned earlier, cluster 1 is the oldest and most undifferentiated group of papers that corresponds with the invention stage. Cluster 4 and 2 represent the emergence of Doxil paradigm, and their intense intercommunication is just the translational process form basic to clinical research. The clusters 3 and 5 are bifurcations of the central paradigm, the former related to targeting and the latter related to hyperthemal therapy. These topological bifurcations could be the novel research context of a new generations of innovations like MCC-46, and ThermoDox. MCC-46 is a inmunoliposomal doxorubicin, developed by Mitsubishi Tanabe Pharma Corp., which is in the first phase of clinical trials [12]. ThermoDox is a thermosensitive liposomal doxorubicin developed by Celsion Corp. in phase III of clinical studies [12].

The emergence of a paradigm implies the self-organizing of a research community. Barenholz, one of the creators of Doxil, takes account of the main actors and institutions that led the innovation process of the first nanodrug [14]. What he says agrees with our results, in the sense that documents with higher centrality and hierarchy of clusters 2 and 4 were conducted by an international collaboration of Barenholz of the Hebrew University-Hadassah Medical School in Jerusalem, Allen of the University of Alberta in Canada, Gabizon of Hadassah University Hospital in Jerusalem and Lasic of Liposome Technology Inc. in Menlo Park, CA (see Figure 1, Tables 2 and 3).

Finally, it is important to keep in mind that convergence of KT, paradigm emergence and innovation are deeply affected by external regulatory processes such as the FDA and patenting processes [15]. Cluster 2, for example, is closely related to a family of patented inventions, described by Barenholz [14], aimed to prolong liposomal lifespan during plasma circulation. Probably the most influent externality affecting the evolution of TI on nanodrugs is the FDA and EMA (European Medician Agency) approval process. From the beginning, Doxil was designed to obtain the approval from the FDA and the EMA [15]. Clinical trials are the main component of the clinical field (clusters 4, 6, 7 and 8), and these must be previously approved in USA by the FDA via the Investigational New Drug (IND) application [15]. Following the clinical trials (phase I, II and III), it is mandatory to summit the New Drug Application to the FDA for a nanodrug enters the market [15]. An analysis of the crosstalk between legal regulatory document, with patent and literature networks for different countries could be fundamental to reaching a deeper understanding on how knowledge translation could be influenced by regulatory processes and intellectual property systems.

## Conclusion

This exploratory research allows to simultaneously map the convergence of KT, TI and paradigm emergence in the first FDA approved nanodrug. In this regard, we have developed a powerful tool for knowledge management in nanomedicine. This work complements with the previous research about KT in liposomes and metallic nanostrutures. This complementation provides us a relevant view of the current stage of cancer nanotechnologies [17]. We have a contrasting view of consolidated and recent types of nanotechnologies, and we have a zoomed out general perspective of cancer nanotechnologies contrasting with a zoomed in view of the paradigmatic successful case of the first FDA approved cancer nanotechnology.

In addition, this methodology has the potential to become a useful evaluating tool for KT. For example, Rajan et al. have pointed out there are two classes of KT models: the T models and the process models [39]. The firsts are focus in describe "phases/components for translational research in different blocks". While, the second type of models describe the process of KT. Rajan et al. state that T models are not appropriated for "evaluating and improving the the performance of translational research". Our research integrates both types of models. This methodology can identify stages of the KT and determinate how strongly- and in a near future how effective and efficient- these stages are interacting with each other.

Much remains to be done; for example, it could be interesting to study how the structure of knowledge is affected when a drug is not approved, how the crosstalk between regulation and research is organized, or which institutions lead the process. These areas of research are currently under way in our group.

**Competing interests**
The author reports no conflicts of interest in this work.

**Author' contributions**
All authors contributed to the research design, interpretation of results and writing of the paper. DF-O built the database, and performed the network analysis and text mining. All authors read and approved the final manuscript.

**Acknowledgments**
David Fajardo is supported by a CONACYT PhD scholarship. The Digital Medical Library of the Faculty of Medicine, UNAM provided access to the Web of Science, which enabled this research.

**Author details**
[1]Graduate program in Medical Sciences and Health, Universidad Nacional Autónoma de México, Mexico City, Mexico. [2]Public Health Department, School of Medicine, National Autonomus University of Mexico (UNAM), Mexico City, Mexico. [3]College of the Souther Border (ECOSUR), Chiapas, Mexico. [4]Centro de Fisica Aplicada y Tecnologia Avanzada, Universidad Nacional Autonoma de Mexico, Queretaro, Mexico. [5]CIATEQ, Queretaro, Mexico.